\documentclass[aps,pra,onecolumn,showpacs]{revtex4}

\usepackage{graphicx, amsmath, placeins}

\begin{document}

\title{Extraction of timelike entanglement from the quantum vacuum}

\author{S. Jay Olson}
 \email{j.olson@physics.uq.edu.au}
 \affiliation{Department of Physics, University of Queensland, St Lucia, Queensland 4072, Australia}
\author{Timothy C. Ralph}
 \affiliation{Department of Physics, University of Queensland, St Lucia, Queensland 4072, Australia}

\date{\today}

\begin{abstract}
Recently, it has been shown that the massless quantum vacuum state contains entanglement between timelike separated regions of spacetime, in addition to the entanglement between the spacelike separated regions usually considered.  Here, we show that timelike entanglement can be extracted from the Minkowski vacuum and converted into ordinary entanglement between two inertial, two-state detectors at the same spatial location --- one coupled to the field in the past and the other coupled to the field in the future.  The procedure used here demonstrates a clear time correlation as a requirement for extraction, e.g. if the past detector was active at a quarter to 12:00, then the future detector must wait to become active at precisely a quarter past 12:00 in order to achieve entanglement.
\end{abstract}

\maketitle
\section{Introduction}
The quantum vacuum is theorized to exhibit a variety of thermal effects associated with spacetime horizons, including Hawking radiation in black hole spacetimes~\cite{hawking1975}, Gibbons-Hawking radiation in de Sitter spacetime~\cite{Gibbons1977}, and Unruh-Davies radiation for accelerated observers in Minkowski spacetime~\cite{unruh1, davies1}. The predicted thermalization of the vacuum in these examples is associated with intrinsic non-classical correlations, i.e. quantum entanglement, between different regions of spacetime. The accessibility of Minkowski spacetime has given rise to research interest in the potential for direct manipulation of this field entanglement~\cite{alsing1, bradler1, mann1, lin1, han1}. However, given that the indirect signature of the entanglement, i.e. the thermalization of the vacuum, has remained well outside the observable regime, the prospects for direct observation of the underlying entanglement have seemed remote.

Very recently, we have predicted a new thermal effect closely related to that of Unruh-Davies, which affects an inertial
particle detector switched on at $t = 0$, having a detector energy gap which is scaled with time as $\frac{E}{at}$ , where $a$ is the scaling constant ~\cite{olson2010a}. Such a detector registers a thermal response to the vacuum, with temperature $T = \frac{\hbar a}{2 \pi k}$. The requirements for observing this effect appear to be much closer to current technological ability than for the Unruh-Davies effect which requires very high accelerations. The origin of the thermalization in this case is that the Minkowski vacuum of massless quantum fields is in fact entangled between timelike separated regions of spacetime, in close analogy to the case of spacelike separation usually considered~\cite{Reznik2005}.

Here we demonstrate the practicality of extracting this entanglement and hence of directly observing the entanglement of the quantum vacuum for the first time. The conceptual novelty of quantum non-separability across time raises some immediate questions about its relation to standard entanglement, which is usually imagined to be a property of a quantum state at a particular time.  We explore the relation to ordinary entanglement by demonstrating that timelike vacuum entanglement can be extracted and converted into ordinary, constant-time entanglement between two detectors, one which couples to the field in the past region P, and another which couples to the field in the future region F of Minkowski spacetime (see Figure 1).  We find that the timelike nature of the entanglement leads to peculiar correlations in time.

\begin{figure}
 \centering
 \includegraphics[scale=0.22]{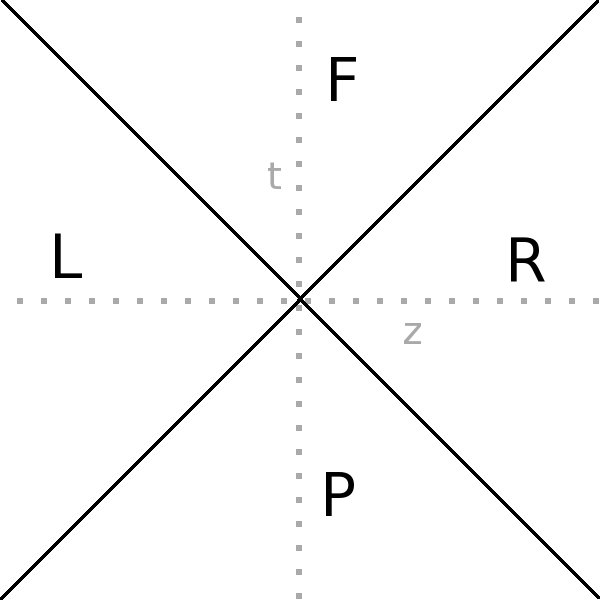}
 \caption{Spacetime divided into quadrants consisting of regions contained by the future and past light cones (F and P), and the right and left Rindler wedges (R and L).}
 \label{fig:  Quadrants of Minkowski spacetime}
\end{figure}

In section 2, we review the specific state of the field in F and P, noting entanglement and the thermal state of the field when restricted to F or P alone.  We also review the thermal single-detector response of the $\frac{E}{at}$ scaled detector.  In section 3, we describe the two-detector entanglement extraction procedure, and a description of the conditions under which entanglement extraction is possible --- the central result of this paper.  Section 4 contains our concluding remarks.

\section{Timelike non-separability in the Minkowski vacuum state}

Before discussing the procedure for extracting timelike entanglement from the vacuum, we first recall the form of the Minkowski vacuum restricted to F-P, and see the entangled state explicitly, for the simplified case of a 1+1 dimensional spacetime.  We examine the left moving sector of a free massless scalar field, for which the left and right moving sectors may be quantized independently, and we use the approximation of a discrete set of modes labeled by $\omega_i$.  

For entanglement to be defined, we require the field in the future to be quantized as an independent system from the field in the past.  For a scalar field $\hat{\phi}(x)$, this is satisfied when $[\hat{\phi}(x_{F}), \hat{\phi}(x_{P}) ]=0$, where $x_{F}$ and $x_{P}$ are any timelike separated points.  This condition is satisfied for massless fields, and in the limit of an arbitrarily small but non-zero mass the commutator is bounded by a maximum value of $\frac{m^2}{8 \pi}$ for any timelike separated points~\cite{greiner1996, bogolyubov1980}, and thus a sufficiently small mass will also allow an approximation of independent systems between $F$ and $P$.

It is well-known that in 2-d spacetime, the Minkowski vaccum restricted to the right, R, and left, L, Rindler wedges (see Fig. 1) can be expressed as an entangled state of the Rindler modes in the following way, which can be understood as the basis for the Unruh effect~\cite{crispino1}:

\begin{align}|0_{M} \rangle = \prod_{i} C_{i} \sum_{n_{i}=0}^{\infty} \frac{e^{- \pi n_{i} \omega_{i} /a}}{n_{i} !}( \hat{a}^{R \dagger}_{\omega_{i}}\hat{a}^{L \dagger}_{\omega_{i}})^{n_{i}} |0_{R} \rangle,
\end{align}
where $|0_{R}\rangle$ is the Rindler vacuum, and $\hat{a}^{R \dagger}_{\omega_{i}}$ is the creation operator for a right Rindler particle, corresponding to the solution $g_{\omega}^{R}(\tau + \epsilon) = (4 \pi \omega)^{-1/2} e^{-i \omega  (\tau + \epsilon)}$ in the coordinate system $t  =  a^{-1} e^{a \epsilon} \sinh(a \tau),  z  =  a^{-1} e^{a \epsilon} \cosh(a \tau)$, with $t$ and $z$ the usual Minkowski coordinates (and analogously for $\hat{a}^{L \dagger}_{\omega_{i}}$).

The Minkowski vacuum restricted to F-P takes an exactly symmetrical form, when expressed in terms of the ``conformal modes'' $g^{F}$ and $g^{P}$~\cite{olson2010a}:

\begin{align}|0_{M} \rangle = \prod_{i} C_{i} \sum_{n_{i}=0}^{\infty} \frac{e^{- \pi n_{i} \omega_{i} /a}}{n_{i} !}( \hat{a}^{F \dagger}_{\omega_{i}}\hat{a}^{P \dagger}_{\omega_{i}})^{n_{i}} |0_{T} \rangle.
\end{align}
Here, F is coordinatized by $ t  =  a^{-1} e^{a \eta} \cosh(a \zeta)$ and $ z  =  a^{-1} e^{a \eta} \sinh(a \zeta)$ with P coordinatized by $ t = - a^{-1} e^{a \bar{\eta}} \cosh(a \bar{\zeta})$ and $ z = - a^{-1} e^{a \bar{\eta}} \sinh( a \bar{\zeta})$.  The creation operators correspond to the solutions $g_{\omega}^{F}(\eta + \zeta) = (4 \pi \omega)^{-1/2} e^{-i \omega (\eta + \zeta)}$ and $g_{\omega}^{P}(\bar{\eta} + \bar{\zeta}) = (4 \pi \omega)^{-1/2} e^{-i \omega (\bar{\eta} + \bar{\zeta})}$, with $|0_T \rangle$ their vacuum.

This symmetry reflects the fact that the conformal modes in $g^F$ are in fact the same solutions as $g^R$, continued from R into F.  

Also mirroring the case of spacelike entanglement, the state of the field in F (or P) alone is a thermal state of the conformal modes:
\begin{eqnarray}
\hat{\rho}_{F} = \prod_{i} \left[ C_{i}^{2} \sum_{n_{i} =0}^{\infty} e^{- 2 \pi n_{i} \omega_{i} /a} |n_{i}^{F} \rangle \langle n_{i}^{F} | \right].
\end{eqnarray}
In 3+1 dimensions, an inertial Unruh-DeWitt detector (that is, one moving along the trajectory $x=y=z=0$, $t= a^{-1} e^{a \eta}$) can also be seen to respond to the vacuum in a manner identical to that of an accelerating detector, provided it is designed to evolve in the conformal time $\eta$, rather than in the proper time $\tau$.  This means that the free Schr\"odinger equation reads:
\begin{eqnarray}
 i \frac{\partial}{\partial \eta} \Psi = H_{0} \Psi 
\end{eqnarray}
where the eigenvalues of $H_{0}$ are taken to have a constant gap, $E$.  Transforming this equation to Minkowski time $t$ reads:
\begin{eqnarray}
 i \frac{\partial}{\partial t} \Psi = \frac{H_{0}}{at} \Psi.
\end{eqnarray}
In other words, we are describing a detector whose energy gap must be scaled as $\frac{1}{at}$ in ordinary Minkowski time $t$.

The interaction term $H_I$ may be taken to be the standard Unruh-DeWitt term, which for a two-state detector takes the form $H_I = \alpha \hat{\phi}(x(t))[|0\rangle \langle 1 | + |1 \rangle \langle 0 |]$.  We thus consider the full Hamiltonian:
\begin{eqnarray}
 i \frac{\partial}{\partial \eta} \Psi = ( H_{0} + e^{a \eta} H_{I} ) \Psi
\end{eqnarray}
The interaction term acquires the exponential factor due to the change of variables to conformal time, and because the coupling to the field is the standard one, and not scaled in time.

From here, one finds the detector reponse function to be:
\begin{eqnarray}
F(E) = \int_{-\infty}^{\infty} d \eta \int_{- \infty}^{\infty} d \eta' e^{-iE(\eta - \eta')} e^{a(\eta + \eta')} D^{+}(\eta, \eta')
\end{eqnarray}
where $D^{+}(\eta, \eta')= \langle 0_{M} | \phi(\eta) \phi(\eta') |0_{M} \rangle.$

The limits of integration correspond to a detector which becomes active at $t=0$.  Another symmetry between the F-P case and the R-L case now becomes important --- the two-point function along the inertial trajectory $x=y=z=0$, $t=a^{-1} e^{a \eta}$ can be calculated to take the form:
\begin{eqnarray}
D^{+}(\eta, \eta') = \frac{a^2 e^{-a(\eta + \eta')}}{4 \sinh^{2}(\frac{a}{2}(\eta - \eta') - i \epsilon)}
\end{eqnarray}
while the two-point function along the trajectory of a uniformly accelerated trajectory $t = a^{-1} \sinh(a \tau )$, $x=y=0$, $z = a^{-1} \cosh(a \tau)$ takes the form:
\begin{eqnarray}
D^{+}(\tau, \tau') = \frac{a^2}{4 \sinh^{2}(\frac{a}{2}(\tau - \tau') - i \epsilon)}.
\end{eqnarray}
This symmetry leads to a formally identical response function integral in the two cases (the additional exponential factor canceling with an exponential factor in the response function integral in the F-P case), and thus to the same thermal signature when the integral is evaluated by standard techniques~\cite{birrell1, crispino1}.  In the case of an accelerated trajectory, the acceleration plays the role of temperature, giving $T_{U} = \frac{\hbar a}{2 \pi c k}$, with one degree Kelvin corresponding to an acceleration on the order of $10^{20} \frac{m}{s^2}$.  In the inertial case, it is the magnitude of the scaling of the detector energy gap, $a$, which plays the role of temperature, giving $T = \frac{\hbar a}{2 \pi k}$, with one degree Kelvin corresponding to a scaling on the order of 100 Gigahertz.  In the following section, we will consider two such scaled detectors --- one in the future and one in the past.

\section{Timelike entanglement extraction}
Here we present our main findings, in three components.  First, we describe the extraction procedure in some detail, and show that extraction of timelike entanglement is possible.  Second, we describe a basic symmetry property of our procedure, namely that identical entanglement can in principle be extracted between regions of arbitrarily great timelike separation, but that larger timelike separations require a correspondingly longer interaction time for the detectors to achieve the same degree of entanglement.  Third, we show a fundamental time correlation in the extraction procedure.

\subsection{Entanglement Extraction Procedure}

We now consider two two-state, energy scaled Unruh-DeWitt dettectors, one of which is active in F, with the other active in P.  Due to the properties of the field commutator (Pauli-Jordan function), for which massless fields satisfies $[ \hat{\phi}(x), \hat{\phi}(y)] = 0$ for timelike separated points $x$ and $y$, the detectors thus interact entirely with independent systems.

In F, the detector moves along the inertial trajectory parameterized by conformal time $\eta$ as $x=y=z=0$, $t = a^{-1}e^{a \eta}$, while in the past, P, the trajectory is parameterized by $x=y=z=0$, $t = - a^{-1}e^{a \bar{\eta}}$.  The detectors will be sensitive to the frequency $E$ with respect to the conformal time in their respective quadrants.  This requires the following energy scaling in terms of the Minkowski time $t$:
\begin{eqnarray}
H_{F} &=& \frac{H_{0}}{at} + H_{I} \\
H_{P} &=& \frac{-H_{0}}{at} + H_{I}
\end{eqnarray}
where $H_{0} |0\rangle = E_{0}|0\rangle$ and $H_{0} |1\rangle= E_{1} |1 \rangle$, with $E_{1} -  E_{0} = E$, which is taken to be positive.  The minus sign appearing in $H_{P}$ cancels with the negative value of $t$ in the past, so that the interpretation of ``ground state'' $|0 \rangle$ and ``excited state'' $|1 \rangle$ is standard for both detectors.

At $t=- \infty$, we take the state of the detectors to be $|00 \rangle$, and we wish to determine the state at $t = \infty$ --- specifically, we will be interested to know whether the state of the detectors is entangled.  Given that the energy gap of the P-detector diverges as it approaches $t=0$, we make the following assumption:  after the P-detector has interacted with the field in P, but before $t=0$, we assume that the detector energy scaling is adiabatically turned off (rather than allowed to ``blow up'' at $t=0$), so that the state of the P-detector is effectively ``frozen'' after its interaction with the field.  Similarly, the energy scaling of the future detector is adiabatically turned on prior to its interaction with the field.

To determine entanglement, we follow an approach analogous to that of Reznik, Retzker, and Silman~\cite{Reznik2005}, who first studied vacuum entanglement extraction from the spacelike separated separated Rindler wedges using two, two-state detectors.  Specifically, we look for a positive value of the negativity of the two-detector state at $t= \infty$, which is the necessary and sufficient condition for the non-separability of the $2 \times 2$ dimensional system formed by our detectors~\cite{Peres1996, Horodecki1996}.

To express the state at $t= \infty$, we use perturbation theory in the conformal time, the same as in the single-detector case, but we now include the ``window functions'' $\chi_{F}(\eta)$ and $\chi_{P}(\bar{\eta})$, defined in F and P respectively, describing the interval over which the detectors are active.

To second order, the state at $t= \infty$ thus takes the following form:
\begin{align}
|\Psi \rangle &= (1 - C)|0_{M} \rangle |00 \rangle \nonumber \\
&- i \int_{-\infty}^{\infty} d \eta \: \chi_{F}({\eta}) e^{a \eta} e^{-i E \eta } \hat{\phi}(\eta) |0_{M}\rangle |01 \rangle \nonumber \\ 
&- i \int_{-\infty}^{\infty} d \bar{\eta} \: \chi_{P}(\bar{\eta}) e^{a \bar{\eta}} e^{i E \bar{\eta}} \hat{\phi}(\bar{\eta}) |0_{M}\rangle |10 \rangle \nonumber \\
&- \int_{- \infty}^{\infty} d \eta \: \int_{- \infty}^{\infty} d \bar{\eta} \: \chi_{F}(\eta) \chi_{P}(\bar{\eta}) e^{a(\eta + \bar{\eta})}e^{-i E(\eta - \bar{\eta})} \hat{\phi}(\eta) \hat{\phi}(\bar{\eta}) |0_{M} \rangle |11 \rangle
\end{align}

To simplify notation, we define the following (unnormalized) field states:
\begin{align}
|A_{F} \rangle &=  \int_{-\infty}^{\infty} d \eta \: \chi_{F}({\eta}) e^{a \eta} e^{-i E \eta } \hat{\phi}(\eta) |0_{M}\rangle \\
|A_{P} \rangle &=  \int_{-\infty}^{\infty} d \bar{\eta} \: \chi_{P}(\bar{\eta}) e^{a \bar{\eta}}e^{i E \bar{\eta}} \hat{\phi}(\bar{\eta}) |0_{M}\rangle \\
|X \rangle &= \int_{- \infty}^{\infty} d \eta \: \int_{- \infty}^{\infty} d \bar{\eta} \: \chi_{F}(\eta) \chi_{P}(\bar{\eta}) e^{a(\eta + \bar{\eta})}e^{-i E(\eta - \bar{\eta})} \hat{\phi}(\eta) \hat{\phi}(\bar{\eta}) |0_{M} \rangle
\end{align}

The state at $t= \infty$ can then be written in the following simplified form:
\begin{align}
|\Psi \rangle &= (1 - C)|0_{M} \rangle |00 \rangle - i |A_{F} \rangle |01 \rangle - i |A_{P} \rangle |10 \rangle - |X \rangle |11 \rangle
\end{align}

We now trace over the field degrees of freedom, and obtain the two-detector density matrix in the basis $|00 \rangle$, $|01\rangle$, $|10\rangle$, $|11\rangle$.
\begin{align}
\rho = \begin{pmatrix}
 N & 0 & 0 & - \langle X | 0_{M} \rangle \\
0 & \langle A_{F} | A_{F} \rangle &  -\langle A_{P} | A_{F} \rangle & 0 \\
0 & - \langle A_{F} | A_{P} \rangle & \langle A_{P} | A_{P} \rangle & 0 \\
- \langle 0_{M} | X \rangle & 0 & 0 & \langle X | X \rangle
\end{pmatrix}
\end{align}
where $ N = 1 - \langle X | X \rangle - \langle A_{P} | A_{P} \rangle - \langle A_{F} | A_{F} \rangle$.

To lowest nontrivial order, the negativity $\mathcal{N}(\rho) $ is given by: 
\begin{align}
\mathcal{N}(\rho) = | \langle 0_{M} | X \rangle | - \sqrt{\langle A_{F}| A_{F} \rangle \langle A_{P} | A_{P} \rangle}
\end{align}

The condition for non-separability of the two-detector state is that $\mathcal{N}(\rho) > 0$.  When the window functions $\chi_{F}$ and $\chi_{P}$ are symmetrical about $t=0$ (as well as the scaling constant $a$ and conformal frequency gap $E$, so that $\langle A_{F}| A_{F} \rangle = \langle A_{P} | A_{P} \rangle$), the non-separability condition amounts to the following:

\begin{align}
\left| \int_{- \infty}^{\infty} d \eta \: \int_{- \infty}^{\infty} d \bar{\eta} \: \chi_{F}(\eta) \chi_{P}(\bar{\eta}) e^{a(\eta + \bar{\eta})}e^{-i E(\eta - \bar{\eta})} \langle 0_{M}| \hat{\phi}(\eta) \hat{\phi}(\bar{\eta}) |0_{M} \rangle \right| \nonumber \\ 
> \left| \int_{- \infty}^{\infty} d \eta \: \int_{- \infty}^{\infty} d \eta' \: \chi_{F}(\eta) \chi_{F}(\eta') e^{a(\eta + \eta')} e^{-i E(\eta - \eta')} \langle 0_{M}| \hat{\phi}(\eta) \hat{\phi}(\eta') |0_{M} \rangle \right|
\end{align}

We have used the notation that $\hat{\phi}(\eta) = \hat{\phi}(x(\eta))$ along the trajectory $t = a^{-1} e^{a \eta}, \vec{x}=0$ in $F$, while in $P$ we have that $t = - a^{-1} e^{a \bar{\eta}}, \vec{x}=0$.

The quantity $\langle 0_{M}| \hat{\phi}(x) \hat{\phi}(x') | 0_{M} \rangle$ takes the ordinary regularized form $- \frac{1}{4 \pi^2} \left[ (t - t' - i \epsilon)^{2} - (\vec{x} - \vec{x}')^{2} \right]^{-1}$ ~\cite{birrell1}, and thus for the given inertial trajectories, a coordinate transformation yields (for appropriately rescaled infinitesimal regulator $\epsilon$):

\begin{align}
\langle 0_{M}| \hat{\phi}(\eta) \hat{\phi}(\eta') |0_{M} \rangle = \frac{-a^{2}e^{-a(\eta + \eta')}}{16 \pi^{2} \sinh^{2}(\frac{a(\eta - \eta')}{2} - i \epsilon) }  \\
\langle 0_{M}| \hat{\phi}(\eta) \hat{\phi}(\bar{\eta}) |0_{M} \rangle = \frac{-a^{2} e^{-a(\eta + \bar{\eta})}}{16 \pi^{2} \cosh^{2}(\frac{a(\eta - \bar{\eta})}{2} - i \epsilon) }
\end{align}

We now consider a specific paired set of window functions:
\begin{align}
 \chi_{F}(\eta) &=  e^{- \eta^{2}}\\
 \chi_{P}(\bar{\eta}) &=  e^{- \bar{\eta}^{2}}.
\end{align}

The entanglement condition $\mathcal{N} > 0$ thus reduces to the following:
\begin{align}
\left| \int_{- \infty}^{\infty} d \eta \: \int_{- \infty}^{\infty} d \bar{\eta} \:  e^{-\eta^{2} - \bar{\eta}^{2}}e^{-i E(\eta - \bar{\eta})} \cosh^{-2}\left(  \frac{a(\eta - \bar{\eta})}{2} \right) \right| \nonumber \\ 
> \left| \int_{- \infty}^{\infty} d \eta \: \int_{- \infty}^{\infty} d \eta' \:  e^{-\eta^{2} - \eta'^{2}} e^{-i E(\eta - \eta')} \sinh^{-2} \left( \frac{a(\eta - \eta')}{2} - i \epsilon \right) \right|.
\end{align}

We express this compactly as $I_X > I_A$.  The first quantity, $I_X$, is not singular, and may be numerically integrated in straightforward fashion with Mathematica.  For $E=1$ and $a=2$, the value of $I_X$ is approximately 1.561.  The second quantity, $I_A$, is formally identical to the (unnormalized) response function integral of an accelerated detector with a window function, where $E$ would represent a fixed proper-energy gap, and the integration variable $\eta$ would represent the proper time (rather than the conformal time it represents here).

To make $I_A$ convenient to compute, we first make use of the identity $\csc^{2}(\pi x) = \frac{1}{\pi^{2}} \sum_{k = -\infty}^{\infty} \frac{1}{(x - k)^{2}}$, so that we can write:

\begin{align}
I_A = &\int_{- \infty}^{\infty} d \eta \: \int_{- \infty}^{\infty} d \eta' \:  e^{-\eta^{2} - \eta'^{2}} e^{-i E(\eta - \eta')} \sinh^{-2} \left( \frac{a(\eta - \eta')}{2} - i \epsilon \right) \nonumber \\
&=  \int_{- \infty}^{\infty} d \eta \: \int_{- \infty}^{\infty} d \eta' \:  e^{-\eta^{2} - \eta'^{2}} e^{-i E(\eta - \eta')} \left( \frac{4}{(a(\eta - \eta') - i \epsilon)^{2}} + \sum_{k = 1}^{\infty}\frac{4}{(a(\eta - \eta') + i \pi k)^{2}} + \frac{4}{(a(\eta - \eta') - i \pi k)^{2}} \right) \nonumber \\
&= I_{\mathrm{inertial}} + \int_{- \infty}^{\infty} d \eta \: \int_{- \infty}^{\infty} d \eta' \:  e^{-\eta^{2} - \eta'^{2}} e^{-i E(\eta - \eta')} \left( \sum_{k = 1}^{\infty}\frac{4}{(a(\eta - \eta') + i \pi k)^{2}} + \frac{4}{(a(\eta - \eta') - i \pi k)^{2}} \right).
\end{align}

The sum/integral in the final line can be evaluated numerically in Mathematica, while the term $I_{\mathrm{inertial}}$ has been evaluated by Satz~\cite{satz2007} with careful attention paid to regularization, since it is formally identical to the response function of an inertial detector with fixed energy gap $E$.  For arbitrary window function, it takes the form:

\begin{align}
 I_{\mathrm{inertial}} &=  \pi E \int_{- \infty}^{\infty} d \eta \: \chi(\eta)^{2} - 2 \int_{- \infty}^{\infty} d \eta \: \chi(\eta) \int_{0}^{\infty} ds \: \chi(\eta - s) \left( \frac{1}{s} - \frac{\cos(E s)}{s^{2}} \right) \nonumber \\
&- 2 \int_{0}^{\infty} \frac{ds}{s^{2}} \int_{- \infty}^{\infty} d \eta \: \chi(\eta) \left[ \chi(\eta) - \chi(\eta - s) \right]
\end{align}
where a change of variables was made to $s = \eta' - \eta''$.  Numerically evaluating this quantity in Mathematica, and combining with the sum/integral term for our chosen window function $\chi_{F}(\eta) = e^{- \eta^{2}}$ yields a value of $I_A$ which is approximately 1.273 (i.e. smaller than $I_X = 1.561$), and thus yields a positive value for the negativity, demonstrating the nonseparability of the detectors at $t=\infty$.

\subsection{Time Translation of the Window Functions}
\FloatBarrier

The value of the negativity calculated in the previous manner is invariant under a simultaneous translation of the window functions in the conformal times $\eta$ and $\bar{\eta}$ by an amount $x$ to the window functions $\chi_{F}(\eta) =  e^{- (\eta - x)^{2}}$ and $\chi_{P}(\bar{\eta}) =  e^{- (\bar{\eta} - x)^{2}}$.  However, different values for $x$ result in different widths of the window functions in the Minkowski time $t$ (see Figure 2).  As the window functions are shifted away from the $t=0$ origin, the detectors must remain active for far longer, in order to achieve the same degree of entanglement.  This supports the intuition that most of the field entanglement is concentrated in the region close to the edges of the light cone.  

\begin{figure}
 \centering
 \includegraphics{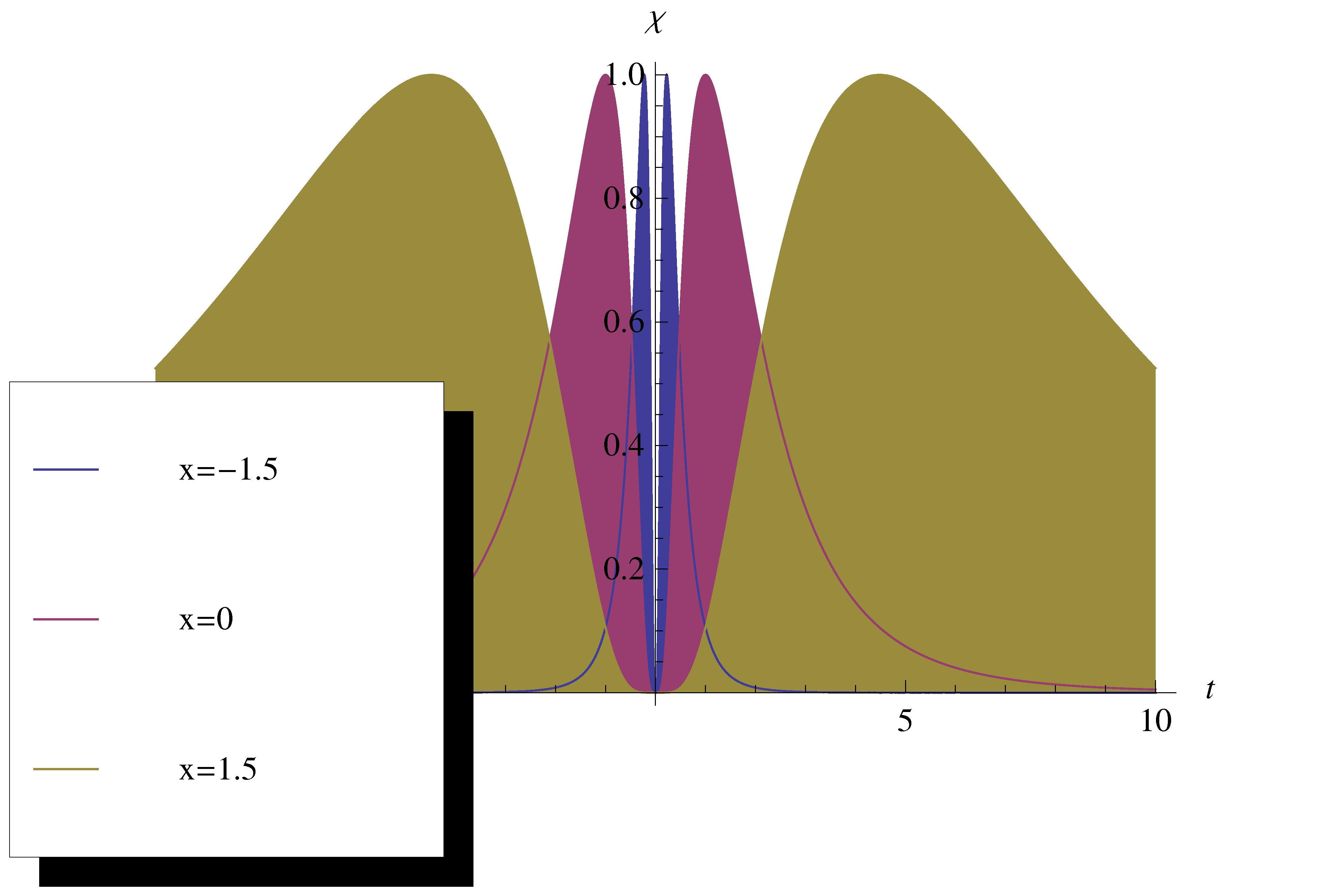}
 \caption{Three symmetrically paired sets of window functions, $\chi_{F}(\eta) = e^{-(\eta - x)^2}$ and $\chi_{P}(\bar{\eta}) = e^{-(\bar{\eta} - x)^2}$, plotted in Minkowski time $t$, which produce identical entanglement between the detectors.  As the window functions are shifted away from $t=0$, the detector must remain active for far longer to achieve the same degree of entanglement.}
\end{figure}

Quantitatively, the price to be paid in terms of the total ``volume'' of interaction time to entangle the detectors to the same degree (as measured by Minkowski time) is exponential in the dispacement $x$ in conformal time.  That is, the ratio $\int_{0}^{\infty} \chi_{F}(\eta - x) dt / \int_{0}^{\infty} \chi_{F}(\eta) dt = e^{ax}$.  This is analogous to the exponential fall-off of entanglement with distance in the spacelike separated case which was observed by Reznik, Retzker, and Silman~\cite{Reznik2005}.

\subsection{Time Correlation}

The negativity does not remain constant, however, if only one of the window functions is shifted, spoiling the symmetry about $t=0$.  In fact, the entanglement can be completely destroyed by shifting one of the window functions sufficiently far away from its symmetrical location in time.  One can readily verify that the quantities $\langle A_F | A_F \rangle$ and $\langle A_P | A_P \rangle$ in equation 18 are each independently invariant under a translation $\chi_F (\eta) \rightarrow \chi_F (\eta - x)$ and $\chi_F (\eta) \rightarrow \chi_F (\eta - x)$, and thus the quantity $I_A$ in equation 24 remains unchanged under the translation of a single window function.  However, the quantity $|\langle 0_M | X \rangle|$ in equation 18 does not possess this symmetry if only one of the two window functions are shifted, and thus the resulting value of $I_X = \left| \int_{- \infty}^{\infty} d \eta \: \int_{- \infty}^{\infty} d \bar{\eta} \:  \chi_{F}(\eta -x) \chi_{P}(\bar{\eta})  e^{-i E(\eta - \bar{\eta})} \cosh^{-2}\left(  \frac{a(\eta - \bar{\eta})}{2} \right) \right|$ will determine whether or not entanglement has been extracted.  We illustrate this by plotting the quantity $I_X - I_A$, which is proportional to the negativity.  The detector state at $t= \infty$ is thus separable only when $I_X - I_A$ is positive.  The window functions plotted are the following:

\begin{align}
 \chi_{F} &= e^{-(\eta - x)^{2}} \nonumber \\
 \chi_{P} &= e^{- \bar{\eta}^{2}}.
\end{align}
  
This corresponds to a situation where the past detector has already interacted with the field at time $\bar{\eta} = 0$, and we must now select the time at which the future detector will be active, with the quantity $x$ signifying how far (in conformal time) we move away from the point of symmetry.  $I_X - I_A$ is plotted as a function of the choice of $x$ in Figure 3 --- positive values correspond to an entangled final state, while negative values correspond to a separable final state.  Clearly, entanglement is maximized around the symmetrical point in time (corresponding to $x=0$), while a sufficiently non-symmetrical choice for $x$ can kill the extraction of timelike entanglement entirely.

\begin{figure}
 \centering
 \includegraphics[scale=1.1]{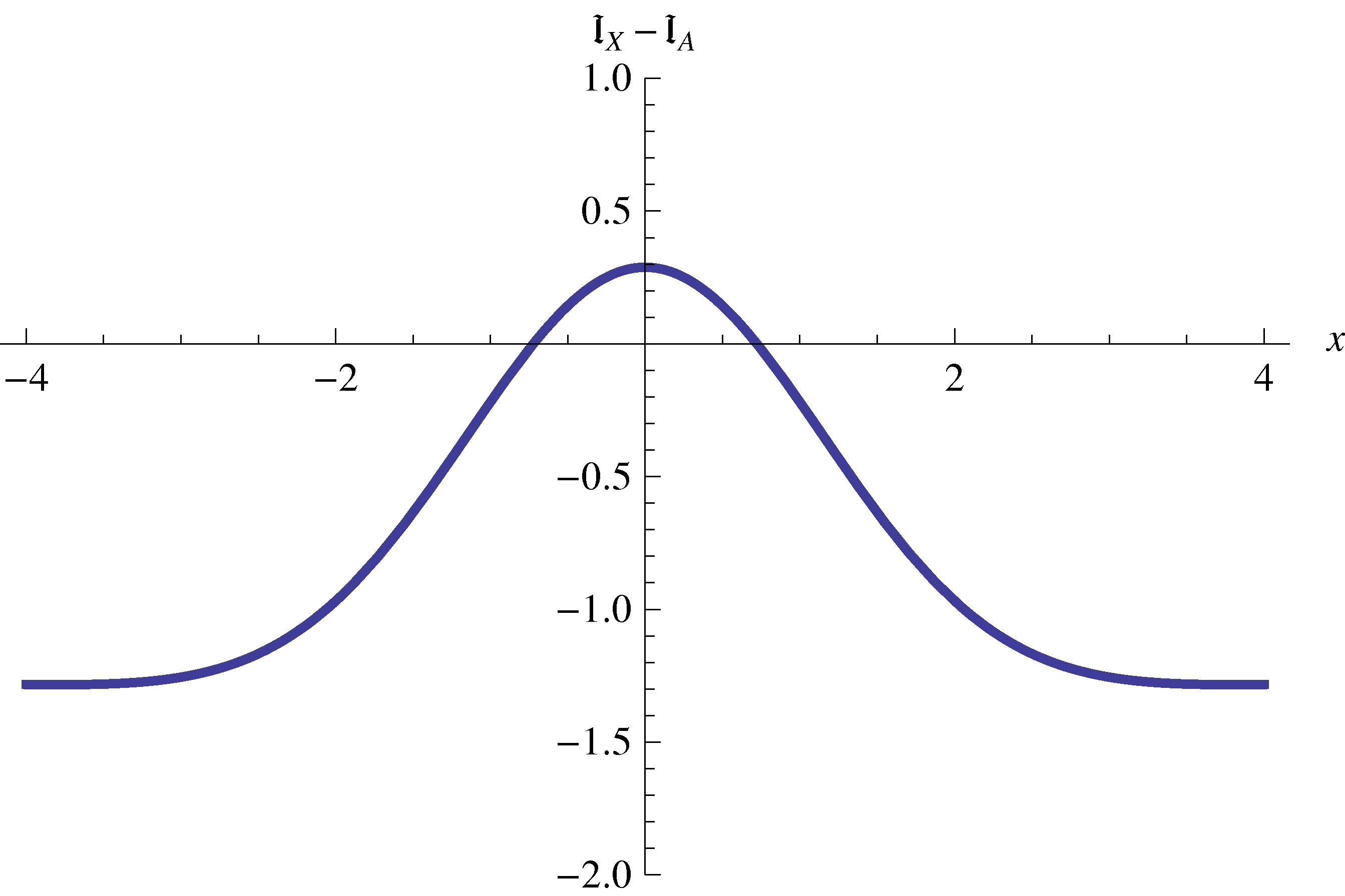}
 \caption{Numerical value of the quantity $I_X - I_A$, for choice of window functions $\chi_{P} = e^{- \bar{\eta}^{2}}$ and $\chi_{F} = e^{-(\eta - x)^{2}}$, for detector parameters $E=1$ and $a=2$.  The detectors are entangled at $t=\infty$ for positive values of  $I_X - I_A$ only.  The position of the window functions must by sufficiently symmetrical about $t=0$ (corresponding to a sufficiently small value of $x$) in order to extract entanglement from the vacuum between the timelike separated regions F and P.}
\end{figure}

Stated more dramatically, a detector which is switched on and off in the vicinity of a quarter to 12:00 can become entangled with a detector interacting with the field at the same spatial location in the future, but only if the later detector waits to be switched on and off at a quarter past 12:00.

\FloatBarrier
\section{Conclusions}

We have attempted to partially answer the question of how timelike entanglement in the Minkowski vacuum can be related to the familiar ``entanglement at a given time'' between simple, two-state systems by showing that timelike entanglement may be extracted and converted into ordinary entanglement between two two-state detectors.  We thus conclude that timelike entanglement may be regarded as a non-classical resource in a manner analogous to the spacelike entanglement that is often studied in the Minkowski vacuum, since any quantum information theoretic protocol may utilize conversion of timelike entanglement to spacelike entanglement as a step in the protocol.

As a thought experiment to illustrate this possibility, we imagine a quantum teleportation protocol in which the entanglement resource is between a detector interacting in P, and a detector interacting in F, and all operations on the P-detector and the qubit-to-be-teleported take place before $t=0$.  Classical information alone is then sent into F, where the F-detector must interact with the field there at a particular time to form the other half of the entanglement resource.  The classical information from P is then used to transform the F-detector into the teleported qubit.  Such a protocol might be called ``teleportation in time,'' since there exists a period after $t=0$ but before the future interaction time where it is not possible to recover the teleported qubit.

We thank Nicolas Menicucci for stimulating discussions.  In addition, we thank the Defence Science and Technology Organisation (DSTO) for their support.

\bibliography{ref2}

\end{document}